\documentclass{pasj00}

\usepackage{natbib} 
\bibpunct{(}{)}{;}{a}{}{,}

\begin{document}
\SetRunningHead{Wedemeyer \& Steiner}{On the plasma flows inside magnetic tornadoes on the Sun}
\Received{2014/02/17}
\Accepted{2014/06/26}

\title{On the plasma flow inside magnetic tornadoes on the Sun}

\author{Sven \textsc{Wedemeyer} %
}
\affil{Institute of Theoretical Astrophysics, 
University of Oslo, Postboks 1029 Blindern, N-0315 Oslo, Norway}
\email{sven.wedemeyer@astro.uio.no}

\author{Oskar \textsc{Steiner}}
\affil{Kiepenheuer-Institut f{\"u}r Sonnenphysik, Sch{\"o}neckstrasse 6, 79104 Freiburg i.Br., Germany}
\affil{Istituto Ricerche Solari Locarno, Via Patocchi 57, 6605 Locarno-Monti, Switzerland}
\email{steiner@kis.uni-freiburg.de}


\maketitle

\begin{abstract}
High-resolution observations with the Swedish 1-m Solar Telescope (SST) and the 
Solar Dynamics Observatory (SDO) reveal rotating magnetic field structures 
that extend from the solar surface into the chromosphere and the corona. 
These so-called magnetic tornadoes are primarily detected as rings or spirals 
of rotating plasma in the Ca II 854.2 nm line core (also known as chromospheric swirls). 
Detailed numerical simulations show that the observed chromospheric plasma motion 
is caused by the rotation of magnetic field structures, which again are driven by 
photospheric vortex flows at their footpoints. 
Under the right conditions, two vortex flow systems are stacked on top of each other. 
We refer to the lower vortex, which extends from the low photosphere into the convection zone,
as intergranular vortex flow (IVF). 
Once a magnetic field structure is co-located with an IVF, the rotation is mediated 
into the upper atmospheric layers and an atmospheric vortex flow (AVF, or magnetic tornado) 
is generated. 
In contrast to the recent work by Shelyag et al., we demonstrate that particle 
trajectories in a simulated magnetic tornado indeed follow spirals and argue that 
the properties of the trajectories decisively depend on the location in the 
atmosphere and the strength of the magnetic field.
\end{abstract}

\section{Introduction}

High-resolution observations with the Swedish 1-m Solar Telescope 
\citep[SST,][]{2003SPIE.4853..341S} in 2008 lead to the discovery of
`chromospheric swirls', which are seen as dark rings or spirals in the 
 core of the Ca\,II infrared triplet line at a wavelength of 854.2\,nm 
 \citep{2009A&A...507L...9W}.  
Subsequent observations with the SST and the Atmospheric Imaging Assembly (AIA) onboard 
the Solar Dynamics Observatory 
\citep[SDO,][]{2012SoPh..275...17L} revealed that these swirls also have an imprint 
in the transition region and corona and that they, at the same time, are connected to magnetic 
bright points in the photosphere below \citep[][ hereafter Paper~I]{2012Natur.486..505W}. 
These observations were explained in Paper~I with the help of numerical simulations, 
which have been carried out with the 3-D radiation magnetohydrodynamics codes 
\mbox{CO$^5$BOLD} \citep{2012JCoPh.231..919F} and Bifrost \citep{2011A&A...531A.154G}. 
The simulations revealed that the observed chromospheric plasma motion is the result of 
rotating magnetic field structures, which are rooted in the top layers of the convection 
zone and extend throughout the atmosphere. 
When the photospheric footpoint and its highly conductive sub-photospheric part 
coincide with a photospheric vortex flow, then such a vortex causes the entire 
field structure to rotate because of the well satisfied frozen-in condition. 
The magnetic field effectively couples the atmospheric layers with each other and 
thus mediates the rotation into the chromosphere, transition region, and corona. 
The visualisation of the simulated velocity field in a single instant of time
produces spiraling streamlines, which remind of terrestrial tornadoes and hence 
led to the name (solar) ``magnetic tornadoes''.  
These events may provide an alternative way of channeling energy and possibly matter 
into the upper solar atmosphere but many details and the resulting net 
contribution to heating the solar corona have yet to be investigated.

We focus here on these small-scale vortex flows in the photosphere and chromosphere and 
do not address rotating magnetic field structures, which are observed on larger 
spatial scales. 
The largest examples, here referred to as ÔgiantÕ solar tornadoes are likely the 
rotating legs of solar prominences
\citep[e.g.,][]{2012ApJ...756L..41S,2012ApJ...752L..22L,2012ApJ...761L..25O,2013ApJ...774..123W}, 
but it is currently not clear if and how they are related to the small-scale events discussed here.

Recently, \citet{2013ApJ...776L...4S} claimed that magnetic field concentrations in the solar 
photosphere do neither produce  magnetic tornadoes nor a ``bath-tub'' effect. 
Here, we demonstrate that, contrary to the claims by  
\citeauthor{2013ApJ...776L...4S}, the simulations presented in Paper~I also show 
spiral-like particle trajectories when considering the full temporal resolution 
of the velocity field. 
We begin with the introduction and definition of different types of vortex flow 
in Sect.~\ref{sec:vortextypes} and analyse particle trajectories based on the 
simulations from Paper~I in Sect.~\ref{sec:parttracks}. 
A discussion of the results and conclusions are presented in Sect.~\ref{sec:discconc}.

\section{Different types of vortex flow} 
\label{sec:vortextypes}

First, we would like to emphasise that there are several types of vortex flows 
that occur on the Sun. 
The formation of a vortex flow in the photosphere is a direct consequence of the 
conservation of angular momentum carried by plasma, which sinks 
down from the surface into the upper convection zone. 
Consequently, a vortex flow is produced, which extends into the upper layers of 
the convection zone. 
This hydrodynamic process is also known as bathtub effect 
\citep[cf. Sect. 3.3 in][]{1985SoPh..100..209N}
because it resembles the swirling water in the sink of a bathtub. 
The effect is most pronounced at the vertices of intergranular lanes, where 
plasma from neighbouring granules converges
\citep[see also, e.g.,][]{2000ApJS..127..159P,2012PhyS...86a8403K}. 
Such flow systems are an integral part of stellar surface convection.
We will refer to this phenomenon as intergranular vortex flow (IVF) hereafter. 
This phenomenon, which is essentially known since the beginnings of hydrodynamic  
simulations of the solar surface layers, has also been called `inverted tornado' by 
\citet{1985SoPh..100..209N}. 
IVFs are also very abundant in the simulations by \citet{2011A&A...533A.126M}
who find vortices with nearly vertical orientation predominantly in intergranular lanes.
They present an example with a horizontal diameter of approximately 80\,km, which 
corresponds to 0.1\,arcsec and would thus be difficult to observe with currently 
available telescopes.  
However, photospheric vortex flows have been observed on different spatial scales and can 
be much larger than an intergranular lane 
\citep[e.g.,][]{1988Natur.335..238B,2008ApJ...687L.131B,2010ApJ...723L.139B,2011MNRAS.416..148V}. 
\citet{2011A&A...533A.126M} suggest that some granular-scale vortices 
could be the peripheral parts of strong but unresolved small-scale vortex flows, i.e. IVFs.

In the presence of magnetic fields, another type of vortex flow can develop in the 
atmospheric layers above an IVF if the foot point of a magnetic field structure 
coincides locally with an IVF. 
The small-scale atmospheric vortex flows, which are produced in this way, have 
been named Ômagnetic tornadoÕ in Paper~I and are referred to as AVFs hereafter in 
order to more clearly distinguish them from IVFs.

\begin{figure}
 \begin{center}
  \includegraphics[width=7.5cm]{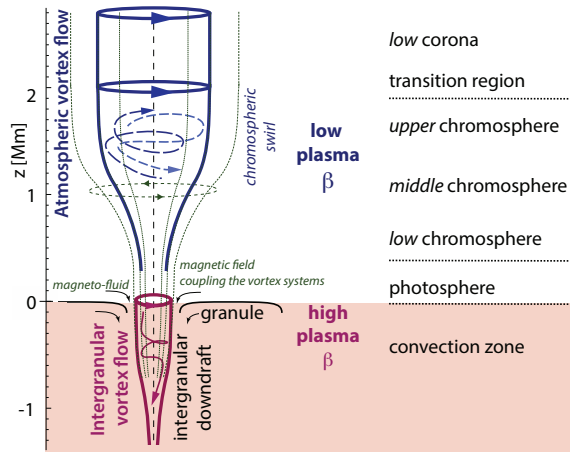} 
 \end{center}
 \vspace*{-3mm}
\caption{Schematic double-nature of vortex flows. 
An atmospheric vortex flow (AVF), also known as `magnetic tornado' as introduced 
in Paper~I can exist on top of a photospheric intergranular vortex flow (IVF), which 
extends into the upper convection zone and drives the AVF above. 
The virtual boundaries of both vortex flows are illustrated with thick solid 
lines. 
In the AVF plasma can propagate upwards and downwards (thin lines with arrows), 
while plasma sinks down in the IVF below. 
The AVF forms under low plasma-$\beta$ conditions, whereas the IVF is 
located in the denser lower layers with a high plasma-$\beta$.  
Both vortex flows are coupled together by the magnetic field (thin line).  
The depicted vortex system is idealised and has a vertically aligned axis
and magnetic field structure, IVF, and AVF being aligned. 
In general, the axis can be tilted and curved, depending on the magnetic field 
at this location. 
}\label{fig:doubletornado}
\end{figure}

The two different types of vortex flow are illustrated in 
Fig.~\ref{fig:doubletornado}. 
Each AVF is driven by an IVF, although IVFs can obviously exist without a 
corresponding AVF. 
The ratio of thermal gas pressure to magnetic pressure, i.e., 
\mbox{plasma-$\beta$}, is large in the upper convection zone of quiet Sun regions. 
There, the gas is significantly ionized so that the magnetic field is essentially frozen-in
and is advected with the convective flows. 
Consquently, the magnetic field is forced to co-rotate with the plasma flow inside an IVF, 
which results in the rotation of the magnetic footpoint and with it in the rotation 
of the entire magnetic field structure.  
A pronounced AVF only develops if the photospheric footpoint of the corresponding 
magnetic field structure is co-located with the IVF long enough. 
As already suggested by \citet{2013JPhCS.440a2005W}, it can happen that the magnetic 
footpoint is moving only temporarily into the IVF or that the IVF is decaying, which 
then results  in only a partial rotation of the magnetic field in the atmosphere and 
thus not in a fully developed AVF. 
A continuous spectrum from partial to full rotation events can be expected. 
In Paper~I, it was shown that these rotating magnetic field structures have detectable 
observational imprints in the chromosphere ('swirls') and in many cases in the corona above. 
Although the detection of chromospheric swirls has been near the observational 
limit so far, it should be possible to derive a more comprehensive picture of 
the full spectrum of swirls regarding, e.g., lifetimes and number of full revolutions 
in the near future with upcoming new, more powerful solar telescopes such as the 
4-m Daniel K. Inouye Solar Telescope 
\citep[DKIST, formerly the Advanced Technology Solar Telescope, ATST][]{2011ASPC..437..319K}
and the 4-m European Solar Telescope \citep[EST][]{2010AN....331..615C}.

\begin{figure*}
 \vspace*{3mm}
 \begin{center}
  \includegraphics[width=16cm]{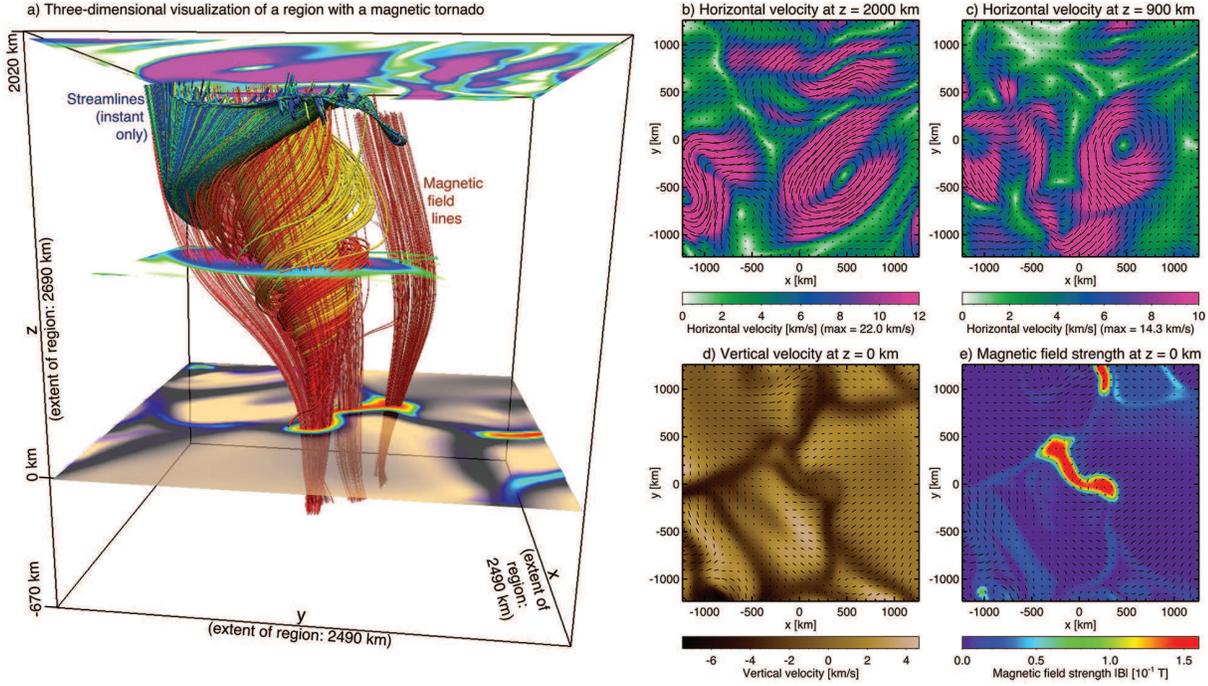}   
 \end{center}
\caption{Visualisation of a close-up region in the numerical model by 
 \citet{2012Natur.486..505W} exhibiting a magnetic tornado. 
The (red) mostly vertical lines in panel~a) represent the magnetic field, 
whereas the spiral lines (red-yellow-green-blue) represent the instant streamlines 
of the plasma in the magnetic tornado. 
The lower plane shows the granulation pattern of the solar surface at a height 
\mbox{$z = 0$\,km} (grey-yellowish shades)  overlaid with the magnetic footpoints 
(blue-green-red), whereas the swirl signature (pink ring) can be seen on the top 
\mbox{($z = 2000$\,km)} and at an intermediate level \mbox{($z = 900$\,km)}.
Please note that the displayed magnetic field lines and streamlines only 
refer to a single instant in time. 
The magnetic field rotates in time and the velocity field evolves, too. 
Consequently, individual (charged) particles do not cross any magnetic field line
but do follow the rotating magnetic field lines. 
The resulting trajectories are shown in Fig.~\ref{fig:parttracks}. 
This image was produced with VAPOR \citep{vapor_clyne2005,vapor_clyne2007}. 
The planes are shown in more detail in panels~b-e: 
b)~ Horizontal velocity at \mbox{$z = 2000$\,km}, 
c)~ horizontal velocity at \mbox{$z = 900$\,km}, 
d)~ vertical velocity at \mbox{$z = 0$\,km} (with negative velocities
for downflows), 
and e)~absolute magnetic field strength at \mbox{$z = 0$\,km}, all 
overlaid with streamlines rendering the horizontal flow field in that instant 
of time. 
}\label{fig:tornadoviz}
\end{figure*}

The distinction between IVFs and AVFs can also be understood in view of the 
different plasma conditions, under which they occur. 
IVFs are produced at the interface between the dense upper convection zone
and the photosphere, where \mbox{plasma-$\beta$} is large. 
Here, the plasma dominates and drags the magnetic field with it. 
In contrast, an AVF forms in the atmosphere in a low \mbox{plasma-$\beta$} environment 
under the influence of a driving IVF below.  
Here, the magnetic field dominates and drags the plasma with it. 
The thick solid lines in Fig.~\ref{fig:doubletornado} illustrate the spatial 
extent of both vortex systems. 
The lines are by no means solid barriers. 
Rather, plasma and associated magnetic field can join and leave the vortex 
system from all sides at all times, which renders the magnetic field structure 
more complicated than can be sketched in the simplified cartoon. 
An impression of this complexity is given in Fig.~\ref{fig:tornadoviz}. 
Inside the vortex flows, plasma particles can propagate downwards and upwards, although 
they are mostly dragged downwards in an IVF.
Being coupled by the rotating magnetic field structure, plasma can propagate between 
the two vortex systems but we find that the flow speeds in the upper photosphere 
are much smaller than in the individual vortex flows above and below.
Rather than of a rigid magnetic flux tube with impenetrable boundaries, one should think
of this flux system as a magneto-fluid in which local magnetic flux concentrations 
are part of a continuous magnetic flux distribution, i.e. part of a continuous magneto-fluid. 
Material from the lateral granular
flow can penetrate and deform the photospheric flux concentration and thus lead to
downdrafts within it.

\citet{2011A&A...533A.126M} provide a further definition and classification of 
vortices based on the so-called `swirling strength' \citep{1999JFM...387..353Z}
and distinguish between mostly vertical and horizontally orientated vortices. 
The vortex flows discussed in this section so far are orientated vertically and 
therefore different from the horizontal vortex tubes reported by 
\citet{2010ApJ...723L.180S}, which 
are found at the edges of granules both in simulations and observations.

Vortex motions in the atmosphere can also be driven by the rapid unwinding 
of twisted magnetic fields instead of the bathtub effect, as it is observed 
in prominence eruptions 
\citep[e.g.,][]{2013ApJ...764...91S,2013AJ....145..153Y}. 
This unwinding process is different from the convective vortex driving 
discussed here.

\section{Particle trajectories} 
\label{sec:parttracks}

\begin{figure}
 \begin{center}
  \includegraphics[width=8cm]{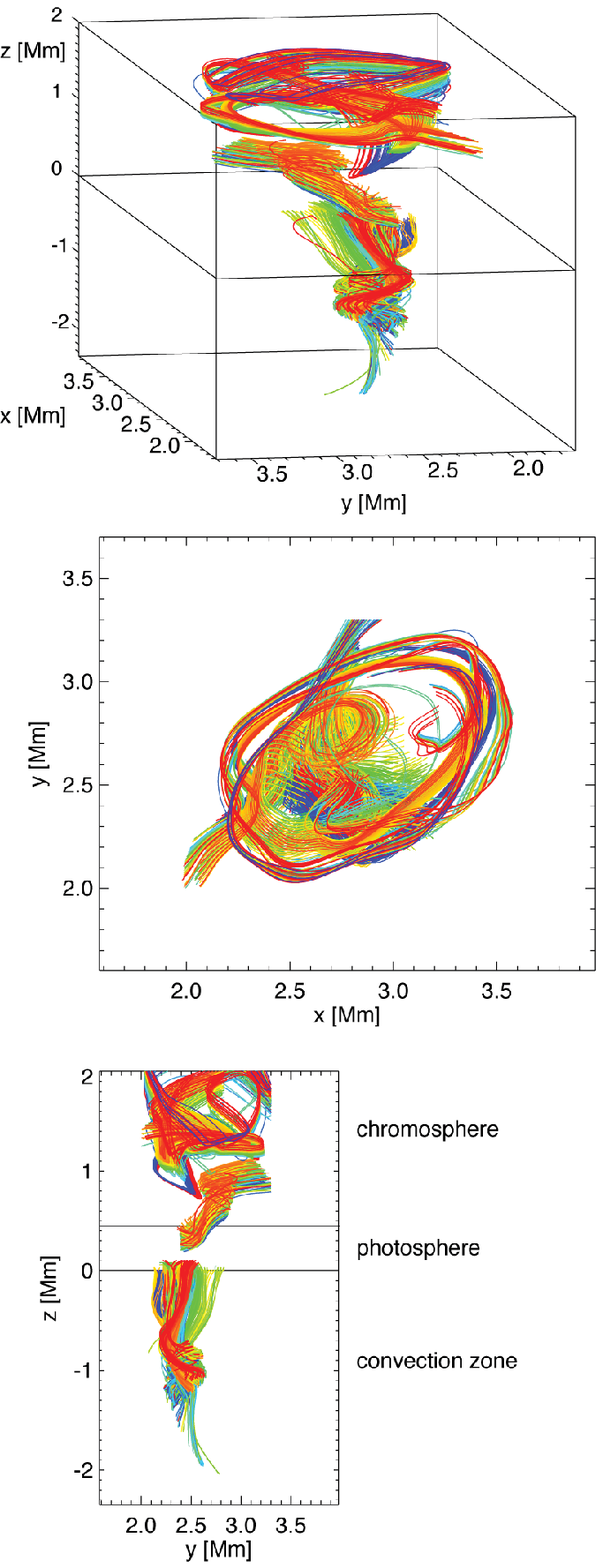}   
 \end{center}
\caption{About 1000~representative particle trajectories in the numerical 
model of the magnetic tornado shown in Fig.~\ref{fig:tornadoviz} 
(see also Paper~I). 
The selected trajectories differ in colour/grey scale and are plotted for 
different perspectives: 
Three-dimensional (top), top view down (middle), and from the side 
(bottom). 
}\label{fig:parttracks}
\end{figure}

The numerical simulation presented in Paper~I consists of a sequence of 
\mbox{CO$^5$BOLD} model snapshots with a cadence of 1\,s, which is part of a much 
longer simulation run. 
In Paper~I, the velocity field is visualized in the form of {\em instant} 
streamlines, which trace the flow field for selected snapshots starting from a 
set of seed points. 
A similar visualisation of the flow field is shown in Fig.~\ref{fig:tornadoviz}. 
These streamlines must not be confused with actual particle trajectories because 
the velocity field evolves while a particle propagates. 
Any particle starting from a point on one of the plotted streamlines would follow 
the streamline in good approximation for a short time so that streamlines still 
give a good overall impression of the propagation of particle ensembles within 
such a flow system.  
The true trajectory of an {\em individual} particle, however, would deviate from 
the initial streamline after a short time. 
True particle trajectories had not been shown in Paper~I but do also exhibit 
spirals, which can be seen here in Fig.~\ref{fig:parttracks}. 
The trajectories are calculated by following a large number of test particles from 
their seed points in each grid cell of the depicted sub-domain through 
the time sequence of snapshots with 1\,s cadence and subsequent refinement to 0.1\,s. 
The considered sequence is  5\,min long. 
For clarity, only about 1000~representative trajectories are plotted.

\begin{table*}
  \caption{Comparison of different numerical simulations, which exhibit vortex 
  flows in the low photosphere. 
  The following radiation magnetohydrodynamics codes are used: 
  \mbox{CO$^5$BOLD} \citep{2012JCoPh.231..919F}, 
  SolarBox \citep{2008ApJ...682.1386J}, 
  MuRAM \citep{2005A&A...429..335V}, 
  and Bifrost \citep{2011A&A...531A.154G}
  }\label{tab:models}
  \begin{center}
    \begin{tabular}{llrrr}
      \hline
Reference&MHD&Initial magnetic&Height of lower&Height of upper\\
         &code&field $|B|_0$   &boundary&boundary\\ 
	  \hline		
Freytag et al. (2012)           &CO$^5$BOLD&  0\,G& -2400\,km& 2000\,km\\
Kitiashvili et al. (2011)       &SolarBox&    0\,G& -5000\,km&  500\,km\\
Kitiashvili et al. (2013)       &SolarBox&    0\,G& -5200\,km& 1000\,km\\
Moll et al. (2012)              &MuRAM&       0\,G&  -900\,km& 800\,km\\
Steiner \& Rezaei (2012)        &CO$^5$BOLD&  0\,G&  -1400\,km& 1400\,km\\
Kitiashvili et al. (2013)       &SolarBox&   10\,G& -5200\,km& 1000\,km\\ 
Carlsson et al. (2010)          &Bifrost&    40\,G& -1400\,km& 14100\,km\\
Steiner \& Rezaei (2012)        &CO$^5$BOLD& 50\,G& -1400\,km& 1400\,km\\
Wedemeyer-B{\"o}hm et al. (2012)&CO$^5$BOLD& 50\,G& -2400\,km& 2000\,km\\
Moll et al. (2012)              &MuRAM&     200\,G&  -900\,km& 800\,km\\   
Shelyag et al. (2013)           &MuRAM&     200\,G&  -800\,km& 600\,km\\
      \hline
    \end{tabular}
  \end{center}
\end{table*}

The three-dimensional visualisation of the particle trajectories again gives  
a tornado-like impression. 
Extremely curved and spiral-like particle tracks can be seen in both the 
chromosphere in the upper part of the domain and in the convection zone below. 
Individual particles, however, may enter and leave the tornado-like region while 
others perform one or several full revolutions around the vortex axis. 
The radius of the particle spirals corresponds to the size of observed chromospheric 
swirls in the upper part of the domain, while the radius is much smaller 
in the convection zone. 
The side view (bottom panel in Fig.~\ref{fig:parttracks}) reveals that essentially no 
particle is travelling the full height range of the tornado during the relatively 
short duration of the considered sequence. 
It appears almost as if the flows above and below the middle photosphere are 
disconnected although both remain of course tightly coupled by the magnetic field. 
Particles, which start from below the middle photosphere, spiral downwards into the 
IVF, while particles, which start above the photosphere, remain in the atmosphere and 
follow the AVF there. 
The particles, which start in the photosphere and below, travel downwards into the 
IVF with absolute velocities $|\vec{v}|$ of up to 10.0\,\mbox{km\,s$^{-1}$} and an 
average of \mbox{$(4.6 \pm 2.0)$\,\mbox{km\,s$^{-1}$}}. 
Particles, which start from positions above the photosphere, can propagate both 
upwards and downwards in the AVF and can also change the vertical direction on the 
way through the vortex.  
The maximum absolute speed $|\vec{v}|$ of the considered particles reaches 
24.5\,\mbox{km\,s$^{-1}$}, with a mean of \mbox{$(13.3 \pm 4.0)$\,\mbox{km\,s$^{-1}$}}.
The corresponding vertical velocity component $v_z$ ranges from -17.0\,\mbox{km\,s$^{-1}$} 
to 19.0\,\mbox{km\,s$^{-1}$}  (where negative velocities indicate downflows).
An AVF would therefore most likely drain material from the chromosphere, which 
possibly results in a net mass transport into the corona.

The {\em apparent} `gap' between the flow systems, which is located in the (upper) 
photosphere, lays in between two different physical regimes: the high 
\mbox{plasma-$\beta$} lower layers and the low \mbox{plasma-$\beta$} upper
layers.
The magnetic field is ``frozen-in'' in both regimes.  
However, while the plasma dominates and advects magnetic field with the convective 
flows in the lower part (i.e., in the IVF) it is vice versa in the upper part, i.e., 
in the AVF. 
There, the magnetic field dominates and is forcing the plasma flows, resulting in the 
observed chromospheric swirls. 
The ionization degree is lowest in the photosphere so that the magnetic field is not 
frozen-in perfectly in the apparent ``gap'', which is nevertheless permeated by the 
rotating magnetic field.  
An IVF is essentially a hydrodynamic phenomenon because it occurs in the high 
\mbox{plasma-$\beta$} regime. 
It can also be (partially) field free and can have a large region of influence from 
where plasma can be dragged into the vortex. 
This way also additional magnetic field can be swept into the vortex and be added to 
the rotating magnetic field structure, which is eventually driving the AVF in the 
chromosphere above.

It should be noted that the magnetic field in the  simulation (mostly vertical 
lines in Fig.~\ref{fig:tornadoviz}, red in the online version) is hardly twisted 
throughout the modelled atmospheric layers, i.e., in the computational domain 
above optical depth unity. 
The model atmosphere extends to only 2\,Mm, which is small compared to the spatial 
scales over which significant twist is to be expected in this low 
\mbox{plasma-$\beta$} regime.

\section{Discussion and conclusions} 
\label{sec:discconc}

Visualisations of the velocity field exhibit clear spiral streamlines in the 
simulations by \citet{2012A&A...541A..68M}, \citet{2012Natur.486..505W},
\citet{2013ApJ...770...37K}, and \citet{2013ApJ...776L...4S}. 
These simulations all started by superimposing a homogeneous vertical magnetic 
field on a relaxed simulation snapshot of solar surface convection but 
differ in terms of spatial extent, initial magnetic field strength and numerical 
details. 
Spiral streamlines are also found in the models by \citet{2011A&A...533A.126M}, 
which were started with weak and initially random magnetic fields. 
Photospheric vortex flows occur also in the hydrodynamic simulations by  
\citet{2011ApJ...727L..50K} and the simulation run v50 by 
\citet{2012ASPC..456....3S} with an initially homogeneous magnetic vertical field with 
\mbox{$|B|_0 = 50$\,G}. 
Swirling motions are observed in the model by \citet{2010MmSAI..81..582C}, which 
was calculated with Bifrost \citep{2011A&A...531A.154G}. 
They suggest that the presented swirl is produced by acoustic waves travelling 
upwards along twisted magnetic field lines
(see also the Supplementary Information in Paper~I).  
Furthermore, \citet{2012A&A...541A..68M} and \citet{2013ApJ...770...37K} provide a 
non-magnetic simulation for comparison. 
The non-magnetic model, from which the simulation by \citet{2012Natur.486..505W} 
was started, is illustrated in \citet{2012JCoPh.231..919F}.  
Some properties of all aforementioned simulations, which are relevant for the 
discussion here, are summarised in Table~\ref{tab:models}.

The models extend to different heights in the atmosphere. 
The model by  \citet{2012Natur.486..505W} includes the upper chromosphere up 
to 2\,Mm, while \citet{2013ApJ...770...37K} put the upper boundary at a height of 
1\,Mm in the middle chromosphere.  
The simulations by \citet{2012A&A...541A..68M}, which were calculated with 
the MURaM code \citep{2005A&A...429..335V}, reach up to a height of 800\,km, i.e. 
into the low chromosphere. 
The model by \citet{2013ApJ...776L...4S}, which was calculated with the MURaM code,  
too, has a comparatively small total height extent of only 
1.4\,Mm. 
Their upper boundary is located at only 600\,km height and thus barely above the photosphere 
and the classical temperature minimum region and lacks the layers where the observable 
imprint of magnetic tornadoes, i.e. chromospheric swirls, is formed.  
For comparison, the onset of shock formation, which is an important structuring agent of the 
chromosphere, only occurs above 700-800\,km \citep[e.g.,][]{1995ApJ...440L..29C,2004A&A...414.1121W}. 
The model by \citet{2013ApJ...776L...4S} thus cannot be used to make conclusions 
about the solar chromosphere and vortex flows therein.

Furthermore, the numerical simulations by \citet{2013ApJ...776L...4S} started with a uniform
vertical magnetic field with a magnetic field strength of $|B|_0 = 200$\,G superimposed on 
a developed non-magnetic model. 
After an initial phase of 40\,min solar time, they write out model snapshots with 
a cadence of 2.41\,s, which are then used for the calculation of trajectories, 
similar to what has been presented in Sect.~\ref{sec:parttracks} of the present paper. 
\citet{2013ApJ...776L...4S} also find swirls in their model, although at much lower 
heights in the solar photosphere. 
However, they report that the corresponding trajectories of test particles based on 
their simulation do not show any spirals and therefore do not have a tornado-like 
appearance. 
\citeauthor{2013ApJ...776L...4S} argue that the particles instead follow  
oscillatory trajectories, for which the magnetic tension in the  
intergranular downflows would play a decisive role. 
They conclude that the horizontal motions in their magnetic field structure are to be 
interpreted as torsional Alfv{\'e}n waves. 
\citeauthor{2013ApJ...776L...4S} finally claim that spiral trajectories (like those
reported in Paper~I) were only present when visualising the velocity field of 
a single model snapshot and that they were absent when considering the 
particle tracks, which incorporate the temporal evolution of the velocity field.

In Sect.~\ref{sec:parttracks}, we demonstrate to the contrary that the simulations 
from Paper~I, which started with an initial magnetic field of $|B|_0 = 50$\,G, do 
show spiral-like particle trajectories when considering the full temporal resolution 
of the velocity field. 
The most pronounced spirals in the streamlines within a single snapshot 
(see Fig.~\ref{fig:tornadoviz}) are seen in the chromospheric part of the magnetic 
field structure, i.e. inside the AVF. 
The model by \citeauthor{2013ApJ...776L...4S} has its upper boundary at only $z = 600$\,km  
and thus does not extend into this layer and can therefore not contain an AVF. 
The spiral trajectories in the simulated AVF appear most clearly at heights 
above $z \sim 700$\,km, i.e. only at chromospheric heights, which are not in included 
in the model by \citet{2013ApJ...776L...4S}. 
The chromospheric origin of AVFs is clear from observations of chromospheric swirls,
which can only be seen in chromospheric diagnostics such as in narrow-band images 
taken in the core of spectral lines of singly ionised calcium. 
However, despite the difference in magnetic field strength, the particle trajectories 
in the model by \citet{2013ApJ...776L...4S} may rather be compared to the particles in 
our IVF as seen in Fig.~\ref{fig:parttracks}. 
There, the trajectories in the upper convection zone are much less extremely wound
than the chromospheric trajectories inside the AVF above. 
Consequently, the particle trajectories in the lower part of the simulation by 
\citet{2012Natur.486..505W} are not in contrast to what is reported by 
\citeauthor{2013ApJ...776L...4S}, while the corresponding streamlines in 
Fig.~\ref{fig:tornadoviz} and particle trajectories in Fig.~\ref{fig:parttracks}
nevertheless exhibit pronounced spiral tracks in the atmosphere above.

The conclusions made by \citet{2013ApJ...776L...4S} can also directly be compared 
to the work by \citet{2012A&A...541A..68M} because both use the MuRAM 
code and use an initial magnetic field with $|B|_0 = 200$\,G. 
\citeauthor{2012A&A...541A..68M} present a counter-rotating pair of vortex flows 
with clear spiral streamlines but unfortunately no particle trajectories. 
However, their simulations also suggest that the formation of vortex flows is an 
integral part of solar convection with important implications for the dynamics, 
structure and heating of the atmospheric layers.

It should be emphasised that the magnetic tornadoes described in Paper~I and 
here are not stationary. 
Instead, the central axis of the flow system, which corresponds to the vortex 
tube cores reported by \citet{2013ApJ...770...37K}, is changing in time, deforms 
and sways back and forth. 
At the same time, particles move up and down in spirals around the central axis. 
The boundary of such a tornado-like flow system is not always sharply defined. 
Particles can enter and leave the ``tornado'' at different heights. 
These effects have to be taken into account for a meaningful visualisation of the 
complicated flow field.

Furthermore, it is plausible that properties like the abundance and strength of 
magnetic tornadoes, i.e. the part visible as chromospheric swirls, depend on the 
magnetic field strength and its topology and can thus be expected to vary between 
different regions on the Sun. 
The bathtub effect is an integral part of non-magnetic or weakly magnetised flows 
in the surface layers of the Sun and other stars but may eventually be 
impeded by magnetic fields above a certain field strength or of complicate topology. 
The correspondingly low \mbox{plasma-$\beta$} conditions may then hinder the plasma flow 
near the solar surface to rotate magnetic foot points. 
A comparison between simulations without and with weak field show that already 
an initial magnetic field of more than $|B|_0 = 10$\,G can change the appearance of the 
chromosphere significantly. 
The otherwise dominant chromospheric shock wave pattern 
\citep[e.g.][]{2000ApJ...541..468S,2004A&A...414.1121W}
is gradually suppressed and  dynamics and structure are instead governed by shear 
and vortex flows in connection with magnetic field concentrations	
\citep[cf.][]{2012A&A...541A..68M,2012ASPC..456....3S,2013ApJ...770...37K}. 
In this sense, the initial magnetic field strength of 200\,G used by 
\citet{2013ApJ...776L...4S} is already high and not representative for 
the regions where chromospheric swirls have been detected on the Sun. 
Therefore we caution any conclusion regarding chromospheric swirls
based on these simulations alone.
Finally, the existence of torsional Alfv{\'e}n waves and their possible role 
for the energy transport within a magnetic tornado had been suggested already 
by \citet{2012Natur.486..505W} and is in no way in contradiction to tornado-like 
flows with spiral particle trajectories.  
A more systematic grid of numerical simulations with different magnetic field 
strengths and corresponding observations is needed to shed light on the detailed 
properties of vortex flows in the solar atmosphere.

\bigskip


The authors acknowledge support by the Research Council of Norway, grants  
221767/F20 and 208026/F50. 

 \ \\

\bibliographystyle{aa} 


%
\end{document}